\documentclass[journal]{IEEEtran}

\usepackage{cite}

\ifCLASSINFOpdf
   \usepackage{graphicx}
\fi
\begin{document}

\title{\huge{Deep Learning based Cephalometric Landmark Identification using Landmark-dependent Multi-scale Patches}}


\author{Chonho Lee, 
        Chihiro Tanikawa, 
        Jae-Yeon Lim 
        and Takashi Yamashiro 
\thanks{Chonho Lee is with Cybermedia Center, Osaka University, Japan. e-mail: leech@cmc.osaka-u.ac.jp.}
\thanks{Chihiro Tanikawa is with Graduate School of Dentistry, Osaka University, Japan. e-mail: ctanika@dent.osaka-u.ac.jp.}
\thanks{Jae-Yeon Lim is with Graduate School of Dentistry, Osaka University, Japan. e-mail: j.lim@dent.osaka-u.ac.jp.}
\thanks{Takashi Yamashiro is with Graduate School of Dentistry, Osaka University, Japan. e-mail: yamashiro@dent.osaka-u.ac.jp.}
}




%



\maketitle

\begin{abstract}
A deep neural network based cephalometric landmark identification model is proposed.
Two neural networks, named patch classification and point estimation, are trained by multi-scale image patches cropped from 935 Cephalograms, whose size and orientation vary based on landmark-dependent criteria examined by orthodontists.
The proposed model identifies both 22 hard and 11 soft tissue landmarks.
In order to evaluate the proposed model, (i) landmark estimation accuracy by Euclidean distance error between true and estimated values, and (ii) success rate that the estimated landmark was located within the corresponding norm using confidence ellipse, are computed.
The proposed model successfully identified hard tissue landmarks within the error range of 1.32$\sim$3.5 mm and with a mean success rate of 96.4\%, and soft tissue landmarks with the error range of 1.16$\sim$4.37 mm and with a mean success rate of 75.2\%.
We verify that considering the landmark-dependent size and orientation of patches helps improve the estimation accuracy.

\end{abstract}

\begin{IEEEkeywords}
orthodontics, medical imaging, landmark identification, deep learning.
\end{IEEEkeywords}

%
\IEEEpeerreviewmaketitle

\section{Introduction}

\IEEEPARstart{O}{rthodontics} usually examine lateral cephalograms, x-ray images taken from the side of the face, and analyze cephalometric landmarks, morphological feature points of hard and soft tissues.
The landmarks are employed to measure the cephalometric components as distance in millimetres and angles in degree~\cite{delaire1997}.
Identifying landmarks is essential in contemporary orthodontic diagnosis and treatment planning.

However, anatomic landmark identification based on the visual assessment of the cephalogram remains a difficult task that requires manual measurements by specially trained clinicians.
Unfortunately, the manual process often leads to measurement errors~\cite{celik2009}.
Such manual identification is considered time-consuming work that takes on average 15 to 20 min from the expert to handle each individual case~\cite{ChenSK2004}. In our case at Osaka University Dental Hospital, a trained clinician takes about 10 to 15 min for identifying about 40 landmarks per individual.
A fully-automated clinical examination of cephalograms would reduce the workload during routine clinical service and would provide orthodontists with more time for optimum treatment planning. 

Various methods to automate landmark identification have been proposed.
%
Image processing approaches~\cite{cardillo1994,rudolph1998,grau01} usually extract contours regarding to landmark locations and identify the landmarks based on the prior-knowledge. Their performance is highly dependent on the image quality, and they cannot find some kinds of landmarks that are not on any characteristic contours. 
Model based approaches~\cite{forsyth1996,saad2006,weining2006} produce models that detect most likely regions including landmarks.
Combined approaches~\cite{kafieh08,keustermans10,vucinic2010} consider active shape/appearance models to refine cephalometric landmarks by combining with template matching.
They have difficulty in selecting the representative template or shape and are not robust to noise and variability in individual.
Machine learning based approaches~\cite{chakrabartty03,daniela05,leonardi09} such as neural network and SVM improve the shape variability and noise tolerance. 
However, these approaches require a large number of cephalograms to improve the accuracy.
Identification performance were not very satisfactory in a practical use.

The recent breakthrough in image recognition technology using deep convolutional neural network (CNN) model~\cite{krizhevsky2012,gu2017} brings dramatic improvement in diagnostic imaging.
This methodology can be used to automatically diagnose the presence of tuberculosis in chest radiographs~\cite{ting2018clinical}, to detect macular degeneration from fundus images~\cite{burlina2017automated}, and to locate malignant melanoma in skin images~\cite{esteva2017dermatologist}.
It is eagerly desired in the field of orthodontics as well, along with the increasing demands for dental healthcare.

\begin{figure}[t]
\centering
\includegraphics[width=3.5in]{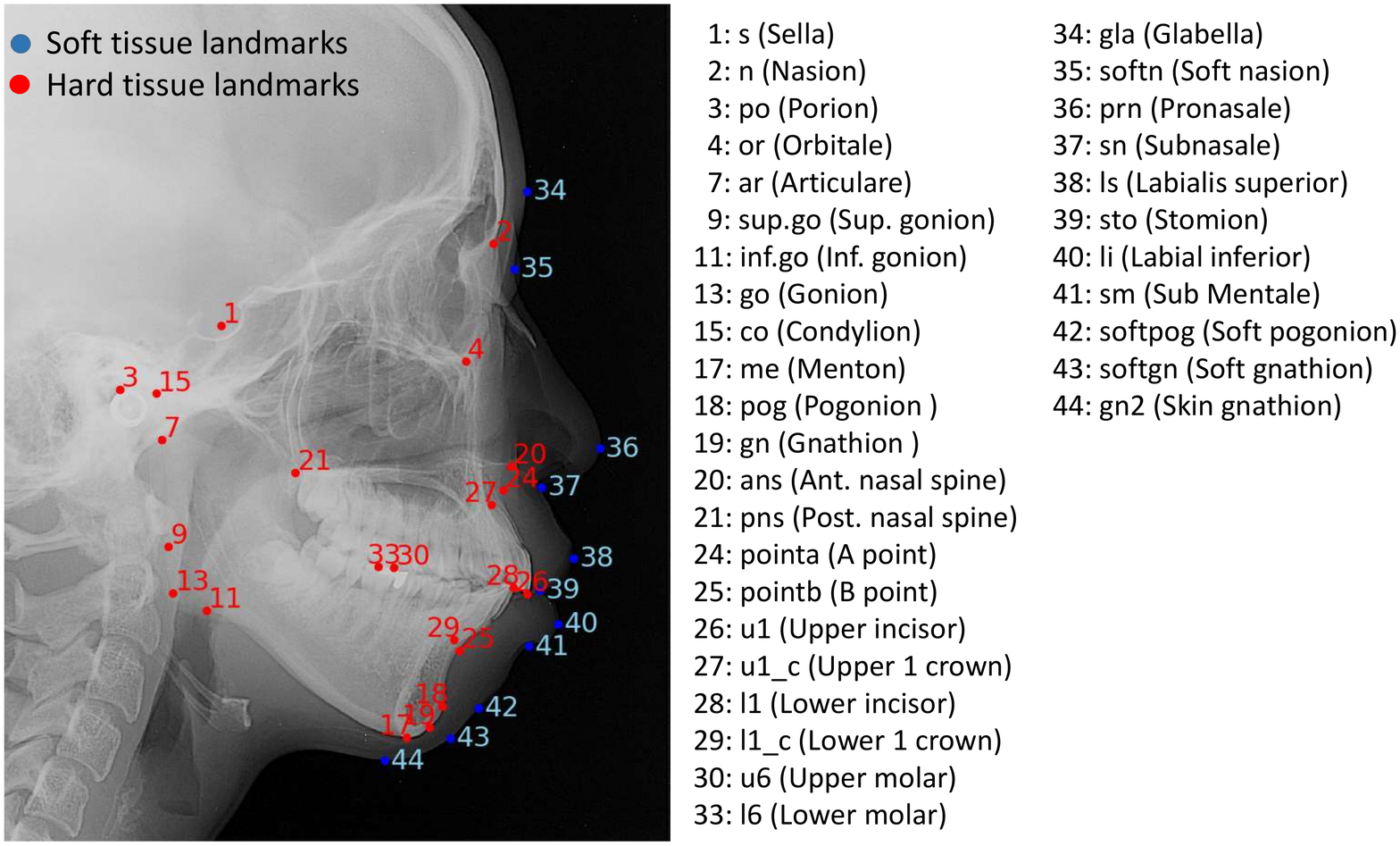}
\caption{A name list of cephalometric landmarks.}
\label{fig:landmarks}
\end{figure}

In this paper, we develop a deep neural network based model for a fully-automated cephalometric landmark identification and demonstrate its feasibility in a practical environment.
Unlike the recently proposed work~\cite{arik17,wang18}, two deep neural networks are trained by multi-scale patches, i.e., cropped images that contain landmarks, whose rectangle size varies based on landmark-dependent criteria examined by orthodontists~\cite{tanikawa2009}.
The proposed model identifies both hard and soft tissue landmarks as shown in Fig.~\ref{fig:landmarks}.

The remainder of this paper is organized as follows.
Section II introduces dataset we prepared and presents the proposed cephalometric landmark identification model.
Section III shows the evaluation results and discusses some future work.

\section{Dataset and Method}
\label{sec:model}

This section presents the proposed cepharometric landmarking model using deep neural networks trained by multi-scale image patches.
The following two subsections explain two different phases, {\it training phase} and {\it landmarking phase}, respectively.
In section \ref{sec:model}-A, we describe two deep neural networks for {\it landmark patch classification} and {\it landmark point estimation}, and how they are trained.
In section \ref{sec:model}-B, we describe how landmarks in a given cephalogram are estimated using the trained neural networks.

\subsection{Training Phase}

\subsubsection{Dataset}

\begin{figure}[t]
\centering
\includegraphics[width=3.5in]{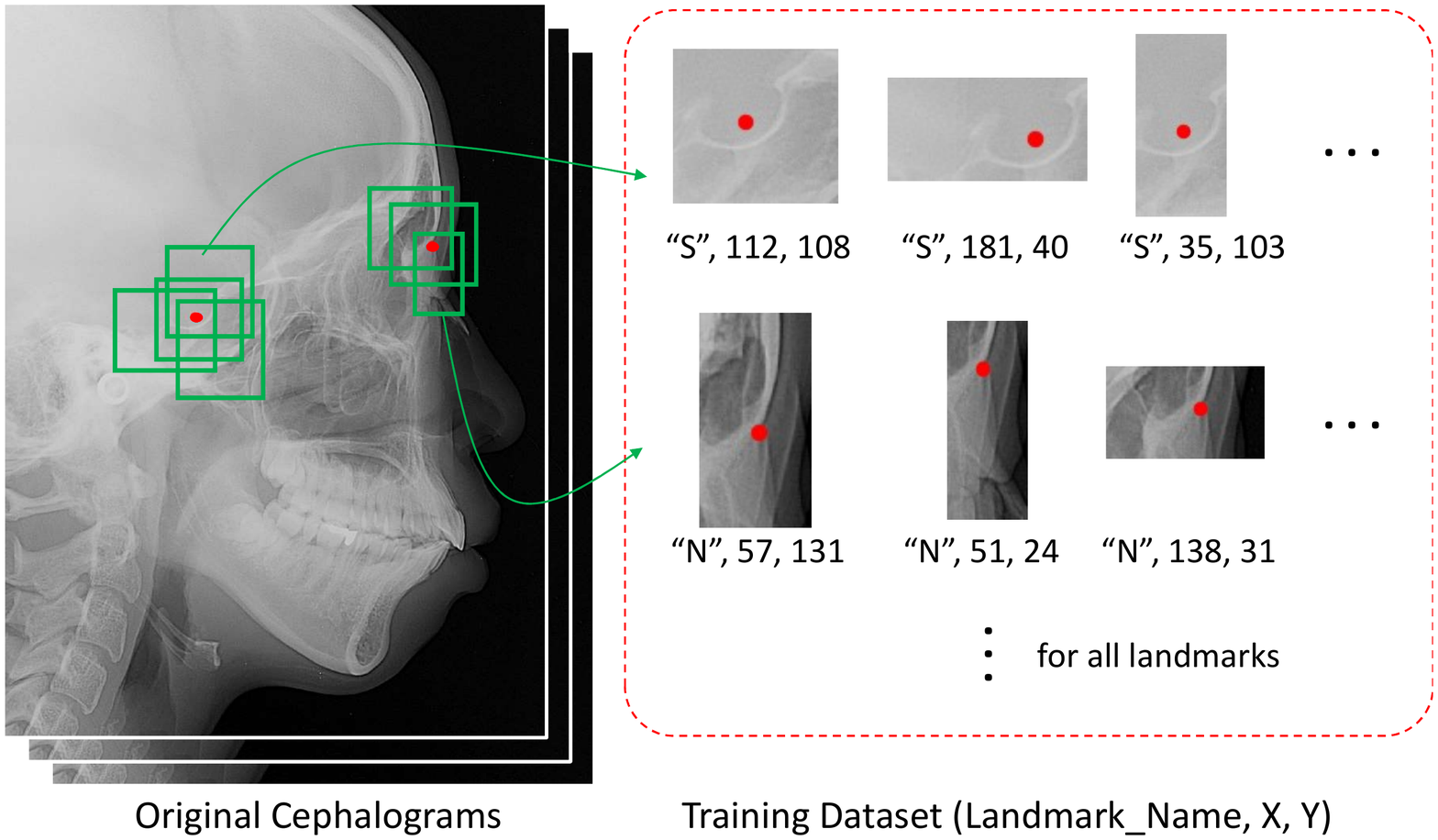}
\vspace{-50pt}
\caption{An example training dataset that contains multi-scale image patches of landmarks with their name and location in the patch.}
\label{fig:trainingdataset}
\end{figure}

A dataset is created based on cephalograms of 936 patients (476 men and 460 women), provided by Osaka University Dental Hospital.
The mean age is 10.98 years $\pm$ 3.57, ranged in 4 and 32 years.
The original cephalogram size is 2100 $\times$ 2500 pixels with a pixel spacing of 0.1 mm. 
As shown in Fig.~\ref{fig:landmarks}, the landmark points (i.e., true values) of 22 hard tissue and 11 soft tissue are plotted by three orthodontists as ground truth values. 
The dataset contains multi-scale image patches including landmarks with their name and location, as illustrated in Fig.~\ref{fig:trainingdataset}.
The landmark name and location are used as labels during the training of neural networks for patch classification and point estimation, respectively.

Patches with various rectangle shapes (green boxes) are extracted, whose width and height range between 80 and 320, and they are resized to 64 $\times$ 64.
The point location is stored as a set of x and y coordinates in the patch, which is not the coordinates in the original cephalogram.
From 935 cephalograms, 9k patches per landmark (i.e., 10 patches per cephalogram) are extracted in total. 

In addition, we perform data augmentation operations to the patches during the training.
For example, we generate patches by a rotation operation with a random angle between -10 to 10 degree, and by a gamma correction with a gamma value between 0.6 and 1.4.

\subsubsection{Structure of deep neural networks}

\begin{figure}[t]
\centering
\includegraphics[width=4.5in]{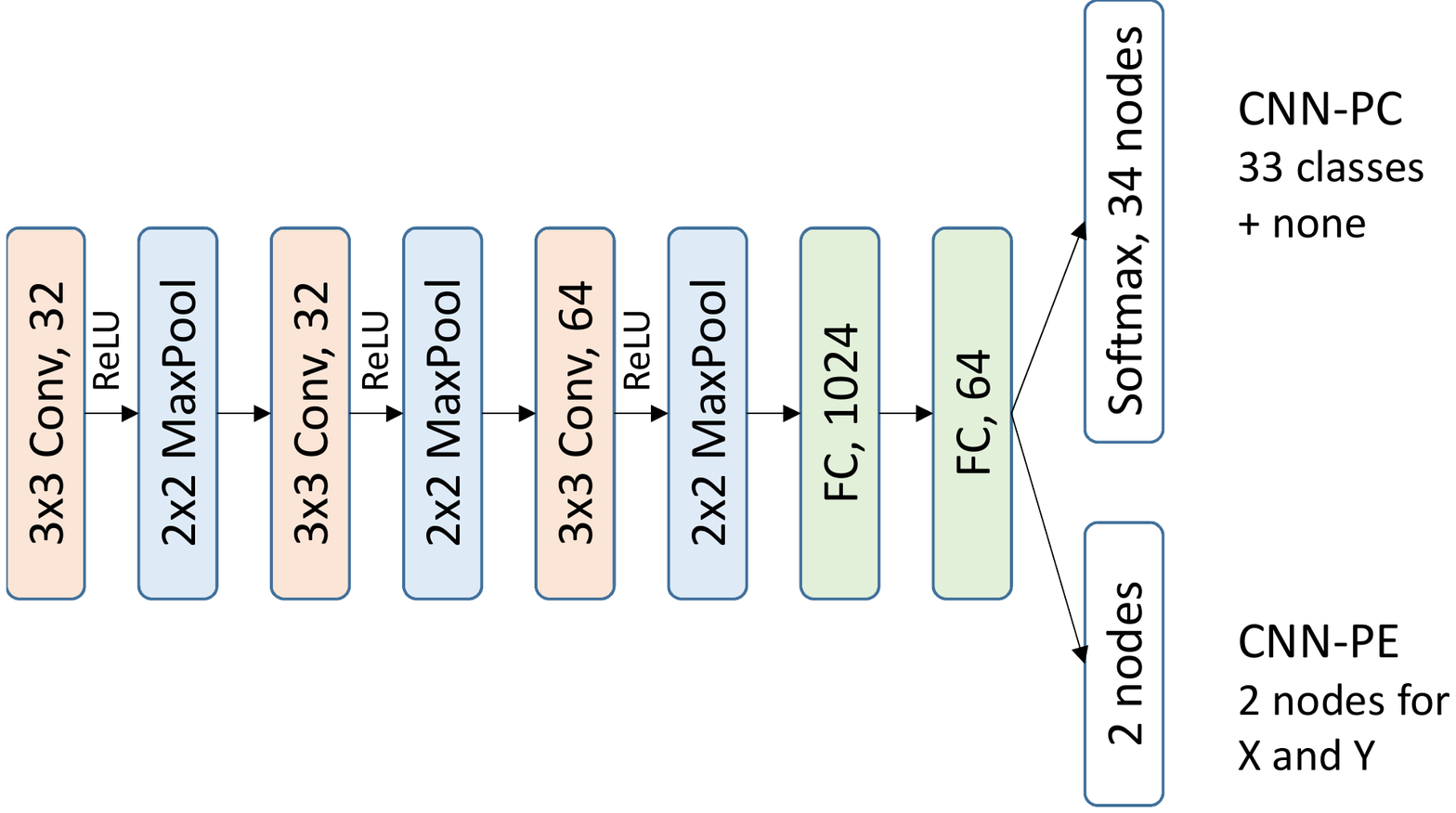}
\vspace{-100pt}
\caption{A structure of CNN-PC and CNN-PE (convolutional neural networks for patch classification and point estimation.)}
\label{fig:cnn_structure}
\end{figure}

\begin{figure}[t]
\centering
\includegraphics[width=3.5in]{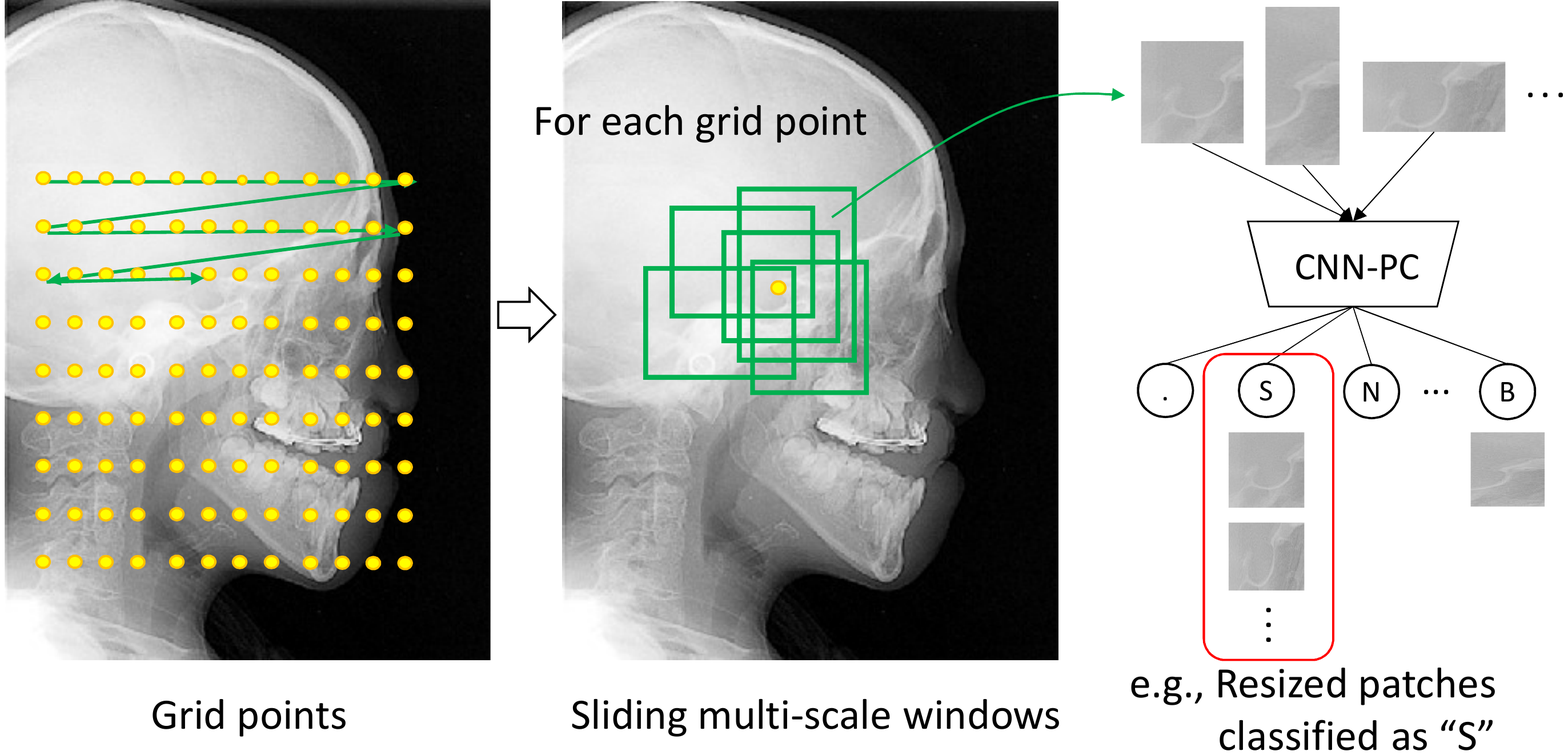}
\vspace{-70pt}
\caption{A flow of patch classification in the landmarking phase.}
\label{fig:patch_classification}
\end{figure}

A convolutional neural network, named CNN-PC, consists of a set of three convolution and pooling layers, and two fully-connected layers with a softmax layer, as shown in Fig.~\ref{fig:cnn_structure}.
Rectified Linear Unit (ReLU) activation is performed after each convolution, and Adam optimizer is used for learning.

Similar to CNN-PC, a convolutional neural network, named CNN-PE, has a two-nodes layer for estimating x and y coordinates, instead of a softmax layer.
In this work, we prepare multiple CNN-PEs, each of which is trained using patches for the corresponding landmark.

\subsection{Landmarking Phase}

In this phase, landmark points for a given cephalogram are estimated using the trained neural networks.
This phase consists of two steps, {\it patch classification} and {\it point estimation}. 

\begin{figure}[t]
\centering
\includegraphics[width=3.7in]{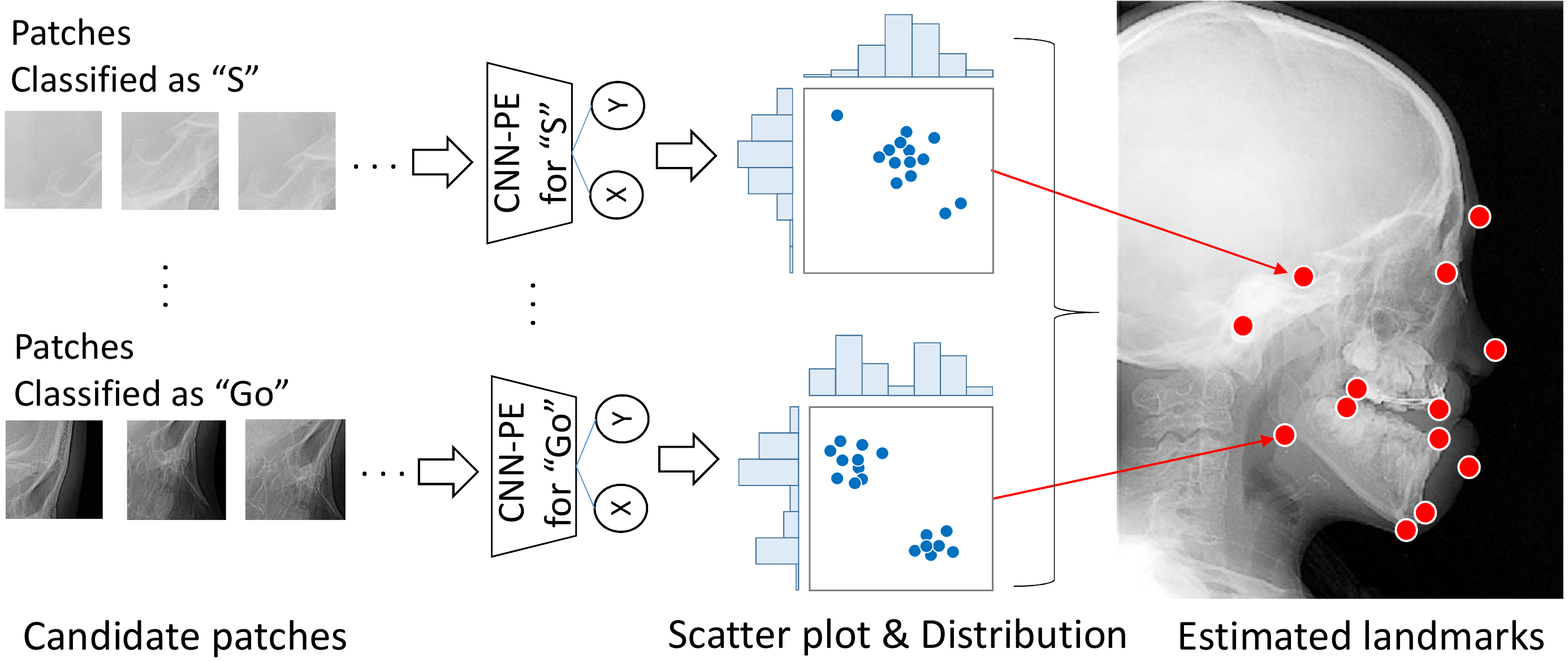}
\vspace{-80pt}
\caption{A flow of point estimation in the landmarking phase.}
\label{fig:landmarking_phase}
\end{figure}

Firstly, given a cephalogram, we set grid points as illustrated in Fig.~\ref{fig:patch_classification}.
For each grid point, we crop multiple patches using multi-scale windows and classify them by CNN-PC.
The classified and resized patches will be stored as {\it candidate patches} of the corresponding landmarks.
Secondly, given a set of candidate patches, x and y coordinates are estimated by the corresponding CNN-PE for each landmark.
An example plot of the estimated coordinates is illustrated in in Fig.~\ref{fig:landmarking_phase} as 2D scatter plot with its distribution.
Based on the scatter plot, we compute one mean/median point as a landmark point.
Before the computation, we omit points as outliers, whose Euclidean distance from the mean/median is greater than two times standard deviation (i.e., $\pm 2 \sigma$).

\begin{figure}[t]
\centering
\includegraphics[width=3.6in]{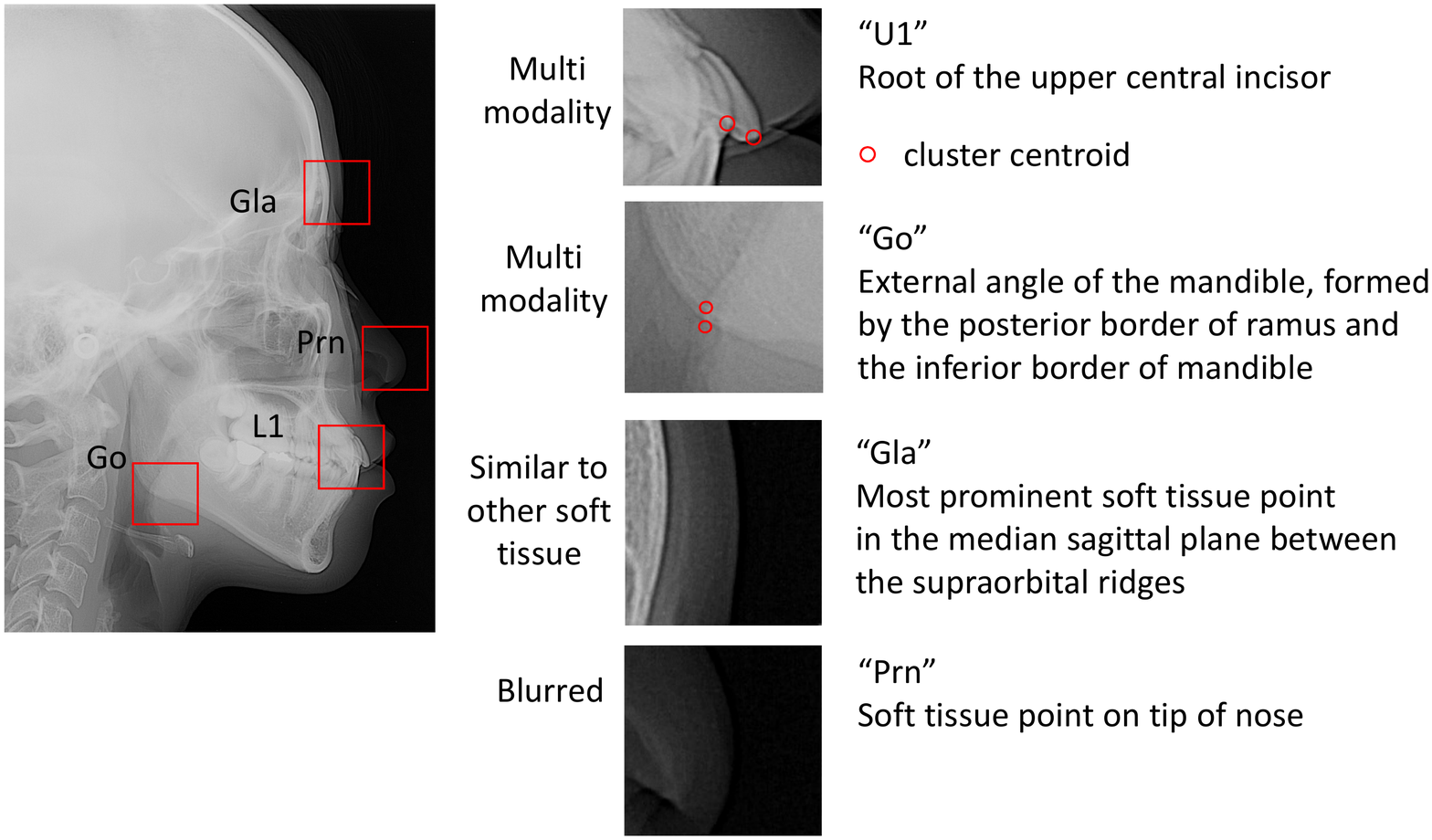}
\vspace{-40pt}
\caption{Examples of ambiguous landmarks (e.g., hard tissues U1 and Go, and soft tissues Gla and Prn) due to multimodality, similarity and blurred image.}
\label{fig:error}
\end{figure}

There are two notes.
One is that the candidate patches may not contain correct landmarks, or are miss-classified due to the performance of CNN-PC.
We later show how the proposed multi-scale patches mitigate the error to improve the landmark estimation accuracy.
Another is that we observed the two cluster distribution for some landmarks as shown in the bottom scatter plot in Fig.~\ref{fig:landmarking_phase}. 
Orthodontists realize this multimodality problem that there are landmarks difficult for them to determine~\cite{tanikawa2009}.
For example, as you can see in Fig.~\ref{fig:error}, there are ambiguous landmarks such as U1 and Go because both right and left side of tooth and jaw are respectively shown in the cephalogram although it is the lateral view.
We later show how the estimation accuracy changes when using the mean or median with/without outliers.

\section{Evaluation}

\begin{figure}[t]
\centering
\includegraphics[width=4.8in]{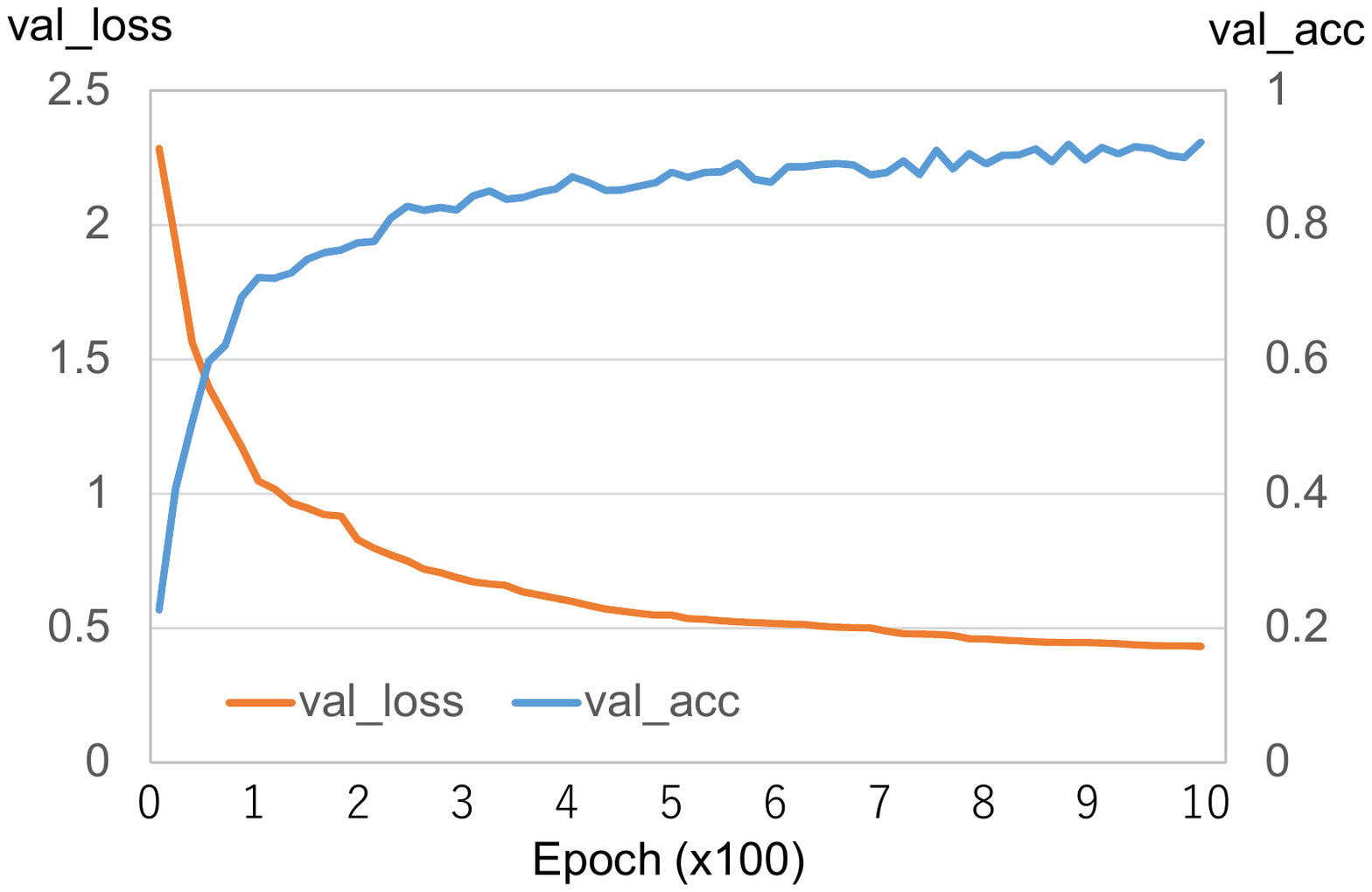}
\vspace{-130pt}
\caption{Validation loss and accuracy of training CNN-PC.}
\label{fig:validation_performance}
\end{figure}

All experiments were performed in Ubuntu 16.04 LTS with a quad-core 3.6GHz CPU, 128GB RAM and Nvidia P100 GPU.
Training the proposed CNN models took 10 sec per epoch for 128 batches.
As shown in Fig.~\ref{fig:validation_performance}, the CNN-PC has achieved around 93\% validation accuracy and 0.4 validation loss in 1k epoch.
For testing, it takes about 0.7 sec to identify one of major landmarks shown in Fig.~\ref{fig:landmarks}. 
In the case of Osaka University Dental Hospital, it dramatically reduces the manual landmarking time that an orthodontist spent about 5 to 7 min to identify 30 major landmarks.

\subsection{Estimation error}

\begin{table}[t]
\renewcommand{\arraystretch}{1.3}
\caption{Comparison results of mean estimation errors (mm) and standard deviations ($\pm$) of hard tissues}
\label{tab:estimation_error}
\centering
\begin{tabular}{|l||c|c|c|}
\hline
         & \multicolumn{3}{c|}{Estimation Error} \\
\hline
Landmark & Saad \cite{saad2006}& Tanikawa \cite{tanikawa2009} & Proposed \\
\hline
1: S  & 3.24$\pm$2.85 & 2.10$\pm$2.52 & 2.06$\pm$1.70  \\
\hline
2: N  & 2.97$\pm$1.85 & 1.70$\pm$1.18 & 1.67$\pm$1.07  \\
\hline
3: Po  & 3.48$\pm$2.46 & 3.63$\pm$2.53 & 2.62$\pm$3.40 \\
\hline
4: Or  & 3.42$\pm$2.43 & 2.24$\pm$1.35 & 2.16$\pm$6.17 \\
\hline
7: Ar  & -- & -- & 1.95$\pm$1.39 \\
\hline
13: Go  & 3.64$\pm$1.76 & 5.06$\pm$3.41 & 3.50$\pm$2.20  \\
\hline
15: Co & -- & -- &  3.50$\pm$4.29 \\
\hline
17: Me  & 4.41$\pm$2.03 & 1.59$\pm$1.07 & 1.35$\pm$1.01  \\
\hline
18: Pog  & 3.66$\pm$1.74 & 1.91$\pm$1.42 & 1.43$\pm$1.32\\
\hline
19: Gn  & 4.22$\pm$1.78 & 1.45$\pm$0.84 & 1.32$\pm$4.16 \\
\hline
20: Ans & 2.70$\pm$1.05 & 2.32$\pm$1.34 & 2.27$\pm$7.22 \\
\hline
21: Pns & -- & -- & 2.33$\pm$3.63 \\
\hline
24: Point a  & 2.55$\pm$0.97 & 2.13$\pm$1.36 & 2.65$\pm$8.34 \\
\hline
25: Point b  & 2.23$\pm$1.24 & 3.12$\pm$9.46 & 2.53$\pm$3.43 \\
\hline
26: U1  & 3.65$\pm$1.59 & 1.78$\pm$2.29 & 2.97$\pm$2.54 \\
\hline
27: U1\_c  & -- & -- & 3.50$\pm$2.85 \\
\hline
28: L1  & 3.15$\pm$2.30 & 1.81$\pm$1.66 & 1.67$\pm$2.90 \\
\hline
29: L1\_c  & -- & -- & 2.30$\pm$2.88 \\
\hline
\end{tabular}
\end{table}

Table~\ref{tab:estimation_error} shows the Euclidean distance error of estimated hard tissue landmarks by the CNN-PE, compared to the ground truth values given by orthodontists.
The results are the average values of 9-cross validation, each of which contains 100 test and 835 training cephalograms, respectively.

The estimation performance for most landmarks improves compared to the previously reported results in Saad~\cite{saad2006} and Tanikawa~\cite{tanikawa2009}, except Point a and U1 that are ambiguous landmarks related to multimodality problem as described in Fig.~\ref{fig:error}.
Besides, the proposed model also estimates additional landmarks such as Ar, Co, Pns, U1\_c and L1\_c, not reported in the previous work.
The proposed model successfully identified landmarks with the range of error between 1.32 mm and 3.5 mm.
89\% of landmarks are located within an range of 3 mm Euclidean distance error.

\subsection{Reliability}

To evaluate the performance reliability, scattergrams that designated errors for manual landmark identification when 10 orthodontists identified a landmark on 10 cephalograms were obtained according to the method reported by Baumrind~\cite{baumrind1971} (for details, see Tanikawa~\cite{tanikawa2009}).
These are employed to assess the performance as the landmark-dependent criteria.
Confidence ellipses with a confidence limit of $\alpha$ were developed for each landmark from the scattergram, using the following equation,
\begin{eqnarray}
\alpha = CHI2 \frac{(x/\sigma_x)^2 - 2\rho(x/\sigma_x \cdot y/\sigma_y) + (y/\sigma_y)^2}{1-\rho^2}
\label{eq1}
\end{eqnarray}
\noindent
where $CHI2$ is the function that provides the one-tailed probability $\alpha$ of the chi-square distribution with 2 degrees of freedom; x and y are the coordinate values; $\sigma_x$ and $\sigma_y$ are standard deviation values for x and y,
respectively; and $\rho$ is the covariance of correlation between x and y~\cite{walpole1978}. The parameter $\alpha$ was assigned to 1.0 when the input position was at the best estimate position, whereas the value was assigned closer to zero when the input was in a biased position from the best estimate position.

\begin{figure}[t]
\centering
\includegraphics[width=3.6in]{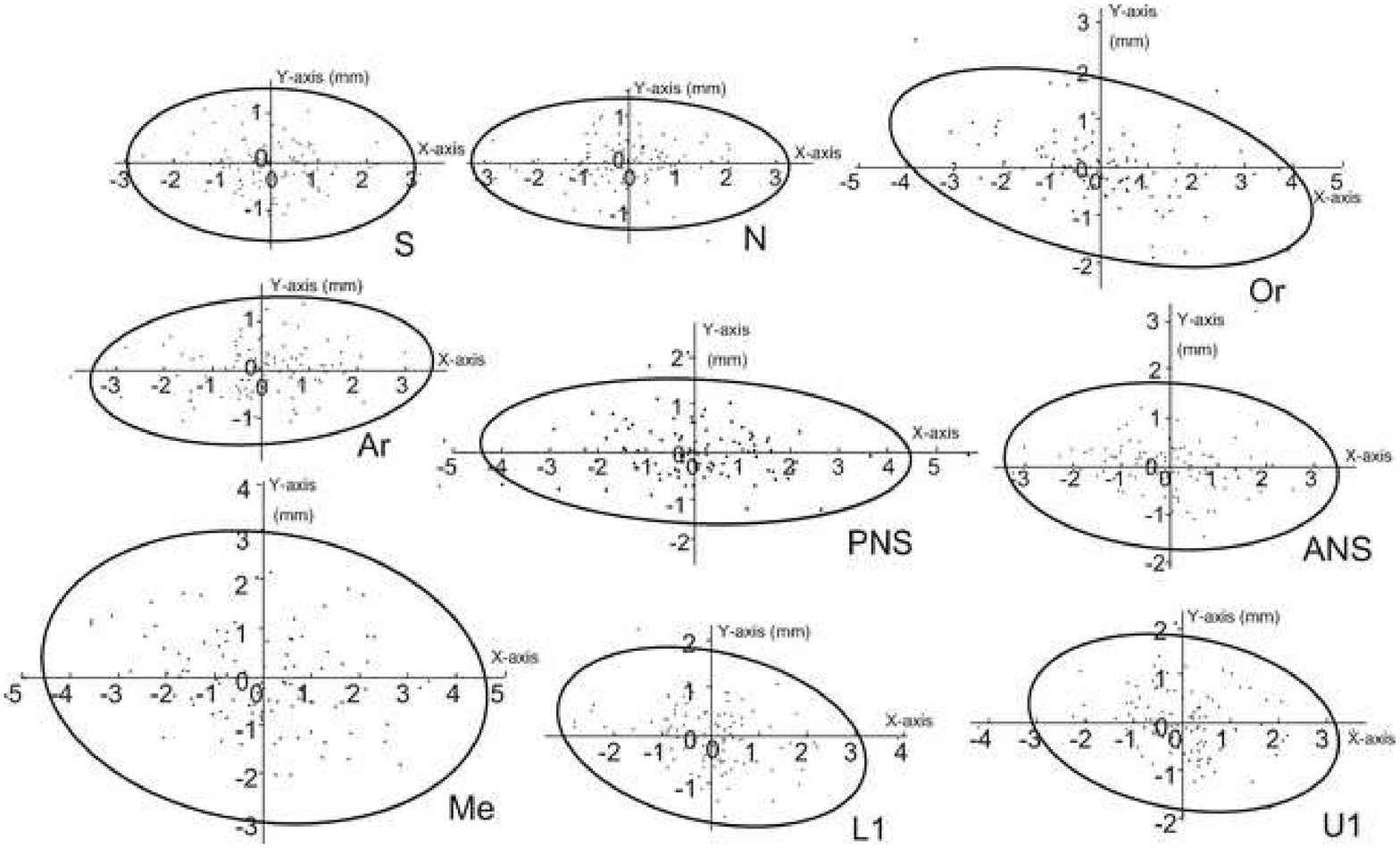}
\vspace{-30pt}
\caption{Example confidence ellipses for obtained landmarks~\cite{baumrind1971} (for more details, see Appendix in Tanikawa~\cite{tanikawa2009}).}
\label{fig:ellipse}
\end{figure}

Example confidence ellipses for obtained landmarks (e.g., S, N, Or, Ar, Pns, Ans, Me, U1, L1) are shown in Fig.~\ref{fig:ellipse}.
Black points indicate coordinate values of landmarks identified by 10 orthodontists. Black lines designate confidence ellipses with $\alpha=0.01$.
The parameters (e.g., the lengths of semimajor and semiminor axes, and angular measurements between x-axis and the semimajor axis) that represent the confidence ellipses are provided in Table~\ref{tab:reliability}.
The reliability performance for all landmarks improves compared to the previously reported results in Tanikawa~\cite{tanikawa2009}.
On average, 96.4\% of estimated landmarks is located within the confidence ellipse.

Finally, we also demonstrate the estimation error and the reliability of soft tissue landmarks in Table~\ref{tab:performance_soft}.
64\% and 91\% of landmarks are estimated within 3 mm and 4 mm, respectively.
As you can see, the reliability performance is much worse than that of hard tissue landmarks.
One of the reason is that patches of most soft tissue images are similar to each other (e.g., Gla and Soft N) or blurred (e.g., Prn and Ls) as described in Fig.~\ref{fig:error}.
We improve the performance by post-processing ignoring outliers based on relative coordinates to other soft tissue landmarks.
Table~\ref{tab:performance_soft} additionally shows the how the post-processing improves the performance.

\begin{table}[t]
\renewcommand{\arraystretch}{1.3}
\caption{Parameters (the lengths of semimajor and semiminor axes, and angular measurements between x-axis and the semimajor axis) that represented the confidence ellipses, and reliability}
\label{tab:reliability}
\centering
\begin{tabular}{|l||c|c|c|c|c|}
\hline
         & \multicolumn{3}{c|}{Confidence Ellipse} & \multicolumn{2}{c|}{Reliability} \\
\hline
         & Angle & Semiminor & Semimajor & Tanikawa & Proposed \\
Landmark & (deg) & Axis (mm) & Axis (mm) & \cite{tanikawa2009} (\%) & (\%) \\
\hline
1: S  & 1.9 & 1.5 & 3.0 & 78 & 91 \\
\hline
2: N  & 1.7 & 1.4 & 3.3 & 77 & 92 \\
\hline
3: Po  & -30.2 & 3.2 & 3.4 & 87 & 97 \\
\hline
4: Or  & 14.1 & 1.8 & 4.2 & 82 & 100\\
\hline
7: Ar  & -4.2 & 1.5 & 3.5 & 82 & 99\\
\hline
13: Go  & 56.1 & 3.7 & 4.7 & 78 & 99 \\
\hline
17: Me  & 9.5 & 3.0 & 4.6 & 100 & 100 \\
\hline
18: Pog  & -63.4 & 3.1 & 5.2 & 95 & 99\\
\hline
19: Gn  & -19.1  & 2.8 & 3.5 & 91 & 99 \\
\hline
20: Ans & 3.7 & 1.7 & 3.5 & 83 & 93 \\
\hline
21: Pns & 2.7 & 1.6 & 4.9 & 80 & 94 \\
\hline
24: Point a  & -71.6 & 3.2 & 3.4 & 88 & 95 \\
\hline
25: Point b  & -62.2 & 3.0 & 4.4 & 98 & 98 \\
\hline
26: U1  & 11.0 & 1.8 & 3.3 & 88 & 91 \\
\hline
27: U1\_c & 68.4 & 3.4 & 6.6 & 94 & 96 \\
\hline
28: L1  & 12.9 & 1.8 & 3.2 & 80 & 96 \\
\hline
29: L1\_c & -49.2 & 2.9 & 6.4 & 94 & 99 \\
\hline
\end{tabular}
\end{table}

\begin{table}[t]
\renewcommand{\arraystretch}{1.3}
\caption{Mean Estimation error and reliability of soft tissue landmarks (with post-processing}.
\label{tab:performance_soft}
\centering
\begin{tabular}{|l||c|c|}
\hline
Landmark & Estimation error  & Reliability (\%) \\
\hline
34: Gla  & 3.58$\pm$5.04 (3.10$\pm$2.04) & 53 (57)  \\ 
\hline
35: Soft N  & 2.13$\pm$6.74 (1.94$\pm$1.65) & 81 (83) \\
\hline
36: Prn  &  3.76$\pm$7.94 (2.31$\pm$2.50) & 52 (66) \\ 
\hline
37: Sn  & 1.16$\pm$2.20 (1.16$\pm$2.20) & 97 (97) \\
\hline
38: Ls  & 5.38$\pm$7.11 (4.37$\pm$2.96) & 41 (51) \\
\hline
39: Sto  & 1.76$\pm$1.55 (1.67$\pm$0.83) & 92 (93) \\
\hline
40: Li  & 2.47$\pm$2.36 (2.26$\pm$1.15) & 76 (78) \\
\hline
41: Sm  & 2.12$\pm$2.49 (1.87$\pm$1.17) & 83 (86)\\
\hline
42: Soft Pog  & 2.21$\pm$5.04 (2.07$\pm$1.69) & 80 (82) \\
\hline
43: Soft Gn & 2.63$\pm$5.86 (2.43$\pm$1.89) & 71 (73) \\
\hline
44: Soft Mn & 3.58$\pm$4.45 (3.31$\pm$1.60) & 59 (61) \\
\hline
\end{tabular}
\end{table}

\section{Conclusion}

We developed a deep neural network based model for a fully-automated cephalometric landmark identification and demonstrated its feasibility in a practical environment.
Two deep neural networks are trained by multi-scale patches, i.e., cropped images that contain landmarks, whose rectangle size varies based on landmark-dependent criteria examined by orthodontists, and 22 hard and 11 soft tissue landmarks are identified.

For evaluation, (i) landmark estimation accuracy by Euclidean distance error between true and estimated values, and (ii) success rate that the estimated landmark was located within the corresponding norm using confidence ellipse, are computed.
The proposed model successfully identified hard tissue landmarks within the error range of 1.32$\sim$3.5 mm and with a mean success rate of 96.4\%, and soft tissue landmarks with the error range of 1.16$\sim$4.37 mm and with a mean success rate of 75.2\%.
For testing, it dramatically reduces the landmarking time from about 5 to 7 min by an orthodontist to 21 sec for 33 landmarks. 

Together with anatomical landmark identification, the diagnosis of anatomical abnormalities using the landmarks is also significant task for orthodontists.
In clinical practice, it is more useful to analyze angles and linear measurements between particular landmarks rather than just point identification.
We will work on evaluating those measurements based on the proposed model in the near future.


%

%



\ifCLASSOPTIONcaptionsoff
  \newpage
\fi



\bibliographystyle{IEEEtran}
\end{document}